# Rapid divergence of the ecdysone receptor in Diptera and Lepidoptera suggests coevolution between ECR and USP-RXR.

François BONNETON[2*], Dominique ZELUS[1*], Thomas IWEMA[2], Marc ROBINSON-RECHAVI[1], Vincent LAUDET[1]

[1] UMR 5665 du CNRS, Ecole Normale Supérieure de Lyon, 46 Allée d'Italie, 69364 Lyon Cedex 07, France

[2] UMR 5534 du CNRS, Université Claude Bernard Lyon 1, Bâtiment Mendel, 16 rue Dubois, 69622 Villeurbanne Cedex, France

* these two authors contributed equally to this work

Corresponding author: François BONNETON

bonneto@univ-lyon1.fr

Phone : (33) (0)4-72-43-13-26

Fax : (33) (0)4-72-44-05-55

**Running Head :** Evolution of ECR and USP-RXR in insects

## ABSTRACT

Ecdysteroid hormones are major regulators in reproduction and development of insects, including larval molts and metamorphosis. The functional ecdysone receptor is a heterodimer of ECR (NR1H1) and USP-RXR (NR2B4), which is the orthologue of vertebrate Retinoid X Receptors (RXR $\alpha, \beta, \gamma$ ). Both proteins belong to the superfamily of nuclear hormone receptors, ligand-dependent transcription factors which share two conserved domains: the DNA-binding domain (DBD) and the ligand-binding domain (LBD). In order to gain further insight into the evolution of metamorphosis and gene regulation by ecdysone in arthropods, we performed a phylogenetic analysis of both partners of the heterodimer ECR/USP-RXR. Overall, 38 USP-RXR and 19 ECR protein sequences, from 33 species, have been used for this analysis. Interestingly, sequence alignments and structural comparisons reveal high divergence rates, for both ECR and USP-RXR, specifically among Diptera and Lepidoptera. The most impressive differences affect the ligand-binding domain of USP-RXR. In addition, ECR sequences show variability in other domains, namely the DNA-binding and the carboxy-terminal F domains. Our data provide the first evidence that ECR and USP-RXR may have coevolved during holometabolous insect diversification, leading to a functional divergence of the ecdysone receptor. These results have general implications on fundamental aspects of insect development, evolution of nuclear receptors, and the design of specific insecticides.

## INTRODUCTION

Ecdysteroid hormones regulate many essential processes in reproduction and development of insects. In *Drosophila*, a single steroid metabolite, 20-hydroxyecdysone (called ecdysone for simplicity), is responsible for controlling the main developmental transitions, including larval molts and metamorphosis (Kozlova and Thummel, 2000). It is a remarkable system in which one simple signal triggers specific transcriptional regulation of several genes, at different stages and in different tissues. Extensive genetic and molecular studies have demonstrated that gene cascades regulated by ecdysone play a central role in the developmental timing in *Drosophila* (Thummel, 2001). Evidence from a few other species supports the conservation of this ecdysteroid regulatory pathway in insects (Henrich and Brown, 1995). But most of these species belong to the very derived holometabolous orders Diptera and Lepidoptera. Insects present a large range of developmental variability, affecting for example ovarian organization (King and Büning, 1985), embryonic germ-band type (Sander, 1976; Patel *et al.*, 1994) or the number of larval molts and the type of metamorphosis (Sehnal *et al.*, 1996; Truman and Riddiford, 1999). Analysis of this diversity at the molecular level is now possible and constitutes a major objective of evolutionary developmental biology.

The functional *Drosophila* ecdysone receptor is a heterodimer of the products of the *ecdysone receptor* (*EcR*) and *ultraspiracle* (*usp*) genes, two nuclear receptors (Koelle *et al.*, 1991; Oro *et al.*, 1992; Yao *et al.*, 1993). Nuclear receptors share a common organization consisting of at least three structural domains: an amino-terminal domain (A/B), a central DNA binding domain (DBD or C domain), and a ligand binding domain (LBD or E domain) (Moras and Gronemeyer, 1998). In addition, a flexible linker region (D domain) is located between DBD and LBD. Some members of this family also contain a carboxy-terminal tail (F domain). The requirement of heterodimerisation between ECR and USP-RXR has been found in other species such as the mosquito *Aedes aegypti* (Wang *et al.*, 2000), the silkmoth *Bombyx mori* (Swevers *et al.*, 1996), and even a member of the Chelicerata, the tick *Amblyomma americanum* (Guo *et al.*, 1998). Understanding the evolution of ecdysone regulation in insects requires comparative analysis of both partners of the heterodimer.

Within the superfamily of nuclear receptors, ECR (NR1H1) belongs to the same group as the vertebrate Liver X Receptors (LXRα and LXRβ: NR1H3 and NR1H2) and Farnesoid X Receptor (FXR : NR1H4) which are also receptors for steroid hormones (oxysterols and bile acids, respectively) (Laudet and Gronemeyer, 2002). Ecdysteroids are not produced by deuterostomes, such as vertebrates. Phylogenies based on 18S rDNA sequences group arthropods and nematodes in the ecdysozoa clade of protostomes sharing the developmental trait of moulting (Aguinaldo *et al.*, 1997). However, ECR horthologues have not been identified in the *C. elegans* genome but only in some parasitic nematodes which are sensitive to ecdysteroids (Sluder and Maina, 2001). Thus, molting regulation and the primary signal are likely to differ among lineages within ecdysozoa. In fact, a recent analysis of more than 100 nuclear proteins does not support the ecdysozoa hypothesis (Blair *et al.*, 2002), and moulting may have appeared several times during metazoans evolution.

USP-RXR (NR2B4) is the orthologue of vertebrate Retinoid X Receptors (RXRα, β, γ : NR2B1, 2, 3) (Laudet and Gronemeyer, 2002). The name USP comes from the phenotype of *Drosophila* mutants (Perrimon *et al.*, 1985), whereas RXR (Retinoid X Receptor) refers to the mammalian ligand (9-*cis* retinoic acid) (Mangelsdorf *et al.*, 1990). In arthropods no mutant phenotype is known outside *Drosophila*, and USP-RXR does not bind 9-*cis* retinoid acid. Now that this gene has been isolated in a wide variety of metazoans, this nomenclature is sometimes confusing in the literature. In this article, we will use the name USP-RXR for all arthropods and simply RXR for orthologues from other taxa. Contrary to ECR, the three-dimensional structure of RXR proteins has been well studied. The crystal structures of the human RXRα LBD (Bourguet *et al.*, 1995; Egea *et al.*, 2000) and DBD (Lee *et al.*, 1993) have been determined, as well as the USP-RXR LBDs of *Drosophila melanogaster* (Clayton *et al.*, 2001) and of the Lepidoptera *Heliothis virescens* (Billas *et al.*, 2001). Comparison of these structures reveals that *Drosophila* and *Heliothis* USP-RXR LBDs are locked in an inactive conformation. Furthermore, authors of these studies suggest that there may be a natural ligand for this USP-RXR, previously seen as an orphan receptor. *In vitro* studies have shown that juvenile hormone III can bind *Drosophila* USP-RXR with a very low affinity (Jones *et al.*, 1997; 2001). This hormone is a sesquiterpenoid chemically analog to retinoids and involved in the control of insect molting and metamorphosis. However, the possibility that juvenile hormone is a natural ligand of USP-RXR awaits further evidence. It has been proposed that arthropods lost the ability to bind 9-*cis* retinoid acid (Escriva *et al.*, 2000). Then this loss may have been followed by acquisition of a new ligand that remains to be identified.

Cloning of ECR or USP-RXR from various arthropods led several authors to observe an intriguing divergence of both proteins in Diptera and Lepidoptera (reviewed in Riddiford *et al.*, 2001). In order to gain further insight into the evolution of ecdysone regulation in arthropods, we performed an evolutionary analysis of both partners. Sequence alignments and structural comparisons reveal a combination of variation and conservation in important functional domains for both ECR and USP-RXR. The major structural divergences are specific to Diptera and Lepidoptera. The most impressive differences affect the LBD domain of USP-RXR. ECR sequences also show variability in other domains, namely the DBD and the carboxy-terminal F domain. Furthermore, we show that the LBDs of both proteins are characterized by an acceleration of divergence rates in the Diptera-Lepidoptera lineage. Our data provide the first evidence that ECR and USP-RXR may have coevolved during the course of holometabolous insect diversification, probably leading to a functional divergence of the ecdysone receptor. They also show that Diptera and Lepidoptera, the most widely used model organisms to analyze ecdysone regulation, are in fact very derived species concerning this developmental system. Therefore, extreme care must be taken when results from *Drosophila* or *Manduca* are generalized, in particular concerning both fundamental aspects of insect development and the design of specific insecticides.

## MATERIALS AND METHODS

*Cloning and sequencing of cDNAs*

New USP-RXR and/or ECR sequences were obtained by RT-PCR from the following species: *Leptopilina heterotoma* (USP-RXR: 850 bp; ECR: 702 bp); *Alfalfa weevi* (USP-RXR: 854 bp); *Periplaneta americana* (USP-RXR: 902 bp); *Folsomia candida* (USP-RXR: 665 bp); *Lithobius forficatus* (USP-RXR: 916 bp) (Table 1).

5 $\mu$g of total RNA were reverse transcribed with random primers and MMLV reverse transcriptase in 20 $\mu$l of reaction mixture according to the manufacturer's instruction (GIBCO-BRL, MMLV-RT kit). The resulting cDNA was amplified by PCR in 100 $\mu$l volume with 10 mM Tris-Hcl pH = 8.3, 50 mM KCl, 1.5 mM $MgCl_2$ (Perkin-Elmer), 0.25mM of each dXTP, 2.5 U Taq Gold DNA polymerase (Perkin-Elmer) and 300 ng of each primer.

Degenerate primers were designed from an alignment of nucleic sequences for either *usp-RXR* or *EcR*. The primers are located within conserved sequences coding the DNA binding and ligand binding domains. Four primers were designed for each gene; their orientation and exact position in *Drosophila* cDNA sequences (*usp*: X53417; *EcR*: M74078) are indicated below into brackets:

usp51: 5' GGI AA(a/g) CA(c/t) TA(c/t) GGI GTI TAC AG (Forward, 499-421)

usp52: 5' TG(c/t) GA(a/g) GGI TG(c/t) AA(a/g) GGI TT(c/t) TT(c/t) AA (Forward, 423-548)

usp32: 5' T(g/t)(c/g) I(g/t)I CGI (c/g)(a/t)(a/g) T(a/g)C TC(c/t) TC (Reverse, 1483-1502)

usp31: 5' GTG TCI CCI ATI AG(c/t) TT(a/g) AA (Reverse, 1597-1616)

ecr51: 5' ATG TG(c/t) (c/t)TI GTI TG(c/t) GGI GA (Forward, 1855-1874)

ecr53: 5' TG(c/t) GAI ATI GA(c/t) AT(c/g) TA(c/t) ATG (Forward, 1984-2004)

ecr33: 5' C(g/t)I GCC A(c/t)I C(g/t)(c/g) A(a/g)C ATC AT (Reverse, 2578-2597)

ecr31: 5’ (c/g)IA (c/t)(a/g)T CCC A(a/g)A (c/t)(c/t)T CIT CIA (a/g)GA A (Reverse, 3001-3025)

For each gene, all combinations of the four primers were used in semi-nested PCR amplifications. Reactions were performed in a Perkin-Elmer Thermal Cycler 480, using a modified “Touch Down” protocol. Briefly, after an initial 10min cycle at 94°C; cycles 1-5: 94°C 1 min, 55°C 1 min, 74°C 2 min; cycles 6-10: 94°C 1 min, 50°C 1 min, 74°C 2 min; cycles 11-15: 94°C 1 min, 45°C 1 min, 74°C 2 min; cycles 16-20: 94°C 1 min, 40°C 1 min, 74°C 2 min; cycle 21-40: 94°C 1 min, 37°C 1 min, 74°C 2 min; followed by terminal elongation for 10 min at 74°C. Extreme care was taken against contamination: PCR were performed in rooms devoted to ancient DNA studies with overpressure, UV lights and dedicated hoods.

PCR products were cloned into a TA cloning vector (Invitrogen) and transformed into competent cells according to the manufacturer’s instructions. Sequencing reactions were performed using a Dye terminator cycle sequencing ready reaction kit with AmpliTaq DNA polymerase FS (Applied Biosystems).

*Protein Sequence analysis*

All available sequences were obtained from NUREBASE (Duarte *et al.*, 2002). Species and accession numbers are shown in Table 1. Protein-coding sequences were aligned using SEAVIEW (Galtier *et al.*, 1996). All positions with gaps were excluded from analyses. Phylogenetic reconstruction was made with Neighbour Joining (Saitou and Nei, 1987) with observed differences as implemented in Phylo_Win (Galtier *et al.*, 1996). The number of complete aligned sites used for tree reconstruction is 74 for ECR DBD, 221 for ECR LBD, 77 for USP-RXR DBD and 145 for USP-RXR LBD. Bootstrap analysis with 1000 replicates was used to assess support for nodes in the tree (Felsenstein, 1985). The phylogenetic tree of RXR/USP sequences is rooted by the jellyfish *Tripedelia cystophora* RXR sequence (Kostrouch *et al.*, 1998). The tree of ECR is rooted by vertebrate LXR and FXR sequences.

Evolutionary distances between sequences were mapped on a pre-defined species consensus tree using Tree-Puzzle (Schmidt *et al.*, 2002), with the JTT substitution model (Jones *et al.*, 1992) plus rate heterogeneity between sites, estimated by a gamma law with eight categories. The consensus tree is based on classical taxonomic data, as well as more specific references concerning the following groups: Diptera (Yeates and Wiegmann, 1999), Lepidoptera (Weller *et al.*, 1992; Regier *et al.*, 2001), Insects (Kristensen 1981; Whiting *et al.*, 1997) and Arthropods (Hwang *et al.*, 2001; Giribet *et al.*, 2001).

In addition, rates were compared between lineages using the relative-rate test on all available sequences (Wilson *et al.*, 1977 ; Robinson*, et al.*, 1998), weighting by the pre-defined tree topology, as implemented in RRTree (Robinson-Rechavi and Huchon, 2000), with a Poisson correction for multiple substitutions.

## RESULTS

### ECR and USP-RXR sequences

In order to study the role of the ecdysone receptor during evolution of arthropod metamorphosis, we analyzed the evolution of its two components: ECR and USP-RXR. When this work was initiated, most of the sequences available in the public databases had been isolated from Diptera and Lepidoptera species. Therefore, it was necessary to investigate a larger sample of insects and other arthropods. Using an RT-PCR approach with degenerated primers located within the DBD and the LBD, we cloned and sequenced cDNA fragments coding for USP-RXR or ECR from five new species (Table 1). These new species give a complete sampling of the different types of metamorphosis in arthropods: holometaboly or complete metamorphosis outside Diptera and Lepidoptera (Hymenoptera and Coleoptera), heterometaboly or incomplete metamorphosis (Dictyoptera), ametaboly or absence of metamorphosis (Collembola), plus one myriapod. Overall, 38 USP-RXR and 19 ECR protein sequences have been used for this analysis, from 33 species. Regarding evolution of these two proteins, as it will be shown further in this article, these 33 species can be distributed into six groups: Diptera (8 species), Lepidoptera (6), other hexapods (7), other arthropods (3), chordates (8) and cnidaria (1) (Table 1). Importantly, the phylogenetic relationships among these six groups are well known and are non ambiguous (see Fig. 1).

### Molecular phylogeny of ECR and USP-RXR

Cloning of ECR or USP-RXR homologues from various arthropods has previously revealed that these proteins are divergent in Diptera and Lepidoptera. This is particularly clear for the LBD of USP-RXR, when sequences from a tick (Guo *et al.*, 1998), a crab (Chung *et al.*, 1998), a locust (Hayward *et al.*, 1999) or a beetle (Nicolaï *et al.*, 2000) are compared to Diptera and Lepidoptera sequences. Although less obvious, the same phenomenon affects ECR (Guo *et al.*, 1997; Saleh *et al.*, 1998; Verras *et al.*, 1999). Understanding this evolutionary divergence should give important insights on the evolution of insect metamorphosis and the functional plasticity of nuclear receptors. Therefore, we performed an analysis of all ECR and USP-RXR sequences together, in order to measure their evolutionary rates and to identify precisely the divergent regions. After sequence alignment, identity percentages and phylogenetic trees were determined separately for the DBD and LBD of both proteins.

Pairwise comparisons show a clear divergence in the LBD of USP-RXR between Diptera-Lepidoptera and other species (Table 2). There is only 49% identity between Diptera-Lepidoptera and other insects, as opposed to 68% between these other insects and other arthropods, and 70% between the other insects and chordates. Thus, the USP-RXR LBD of many insects is less similar to Diptera and Lepidoptera than it is to the chordate RXRs. The same is true of the DBD and LBD domains of ECR (Table 2), although the divergence is less pronounced.

A Neighbour Joining analysis with observed differences performed using the full-length LBD of ECR or USP-RXR obtained the trees shown in Figure 2. Similar topologies are found by parsimony analysis (data not shown; see also: Guo *et al.*, 1997; Guo *et al.*, 1998; Hayward *et al.*, 1999). It should be emphasized that our aim is not to reconstruct the phylogeny of species using RXR or ECR as markers, but rather to characterize the evolution of these receptors using phylogeny as a tool. The trees are therefore presented here to illustrate the aberrant topology with regard to insect phylogeny, and to show the length of the branches. In both trees, it can be seen that Diptera and Lepidoptera sequences constitute a monophyletic group separated from all the other insects. The bootstrap score for the branch that separates Diptera-Lepidoptera from other insects is 100% (boxed). All the other insects are grouped in a branch with a high bootstrap value:100% for ECR (Fig. 2A) and 87% for USP-RXR (Fig. 2B). These topologies are clearly in contradiction with the phylogeny of the species (Fig. 1). For example, the coleoptera *Tenebrio molitor* belongs to the holometabolous insects, a monophyletic group which includes Diptera and Lepidoptera. However the USP-RXR LBD from this beetle is more similar to that of a chelicerate (*Amblyomma americanum*), or even of a chordate, than of a Diptera such as *Drosophila*. The trees of Figure 2 also show that Diptera and Lepidoptera sequences share long branches, when compared to other arthropods proteins. This observation is indicative of a rapid rate of divergence.

**Analysis of evolutionary rates**

The phylogenetic analysis suggests that USP-RXR and ECR sequences have undergone accelerated evolution in the Diptera-Lepidoptera lineage. We therefore decided to estimate and to compare the rates of divergence between groups of species.

In order to obtain the best estimates of branch lengths, we used a constraint topology based on the known phylogenetic relationships between all the species analyzed in this article (Fig. 1). Evolutionary distances between sequences were mapped on this pre-defined species consensus phylogeny. The trees obtained by this method are shown in Figure 3. Moreover, rates were compared between lineages using the relative-rate test on all available sequences. Results are shown in Table 3, as differences of substitution rate between groups of species. From these analyses, it appears that both ECR and USP-RXR LBD sequences of Diptera and Lepidoptera have evolved at significantly different rates than other species (Fig. 3B, C and Table 3). The strongest rate difference is with USP-RXR LBDs. Despite the important distances obtained by mapping ECR DBD sequences on the pre-defined tree for Diptera-Lepidoptera (Fig. 3A), rate differences are not significant for DBDs (data not shown). This may be due to the small numbers of sites available for the test (80 amino-acids).

Our data clearly show that both ECR and USP-RXR experienced a very strong acceleration of evolutionary rate in Diptera and Lepidoptera versus other insects. It is therefore essential to identify which regions of the proteins were affected by this acceleration.

**Divergence of the Ligand Binding Domain of USP-RXR**

It has been shown recently that both crystal structures of a Lepidopteran (*Heliothis*) and a Dipteran (*Drosophila*) USP-RXR LBD are locked in an unusual antagonist conformation (Billas *et al.*, 2001 ; Clayton *et al.*, 2001). Sequence alignment clearly shows the differences between the LBD of USP-RXR proteins from Diptera and Lepidoptera and their homologues in other arthropods. They are grouped into three divergent domains and are not located randomly along the sequence (Fig. 4).

Interestingly, many differences affect precisely two regions that are implicated in the unusual conformation of *Drosophila* and *Heliothis* USP-RXRs: the loop between helices H1 and H3, and the carboxy-terminal end of the LBD (helix H12 and the loop between H11 and H12) (Fig. 3). Helix H12 is locked in an inactive position by making contacts with the loop H1-H3, specifically with a conserved domain of 13 residues (boxed in gray in Fig. 4). This domain is well conserved within the lineage of Diptera and Lepidoptera, but is absent in other arthropods, where the loop H1-H3 is highly variable in length and in sequence. Furthermore, three (*Heliothis*) or four (*Drosophila*) residues of the conserved region interact with the phospholipid ligand cocrystallized with the LBD (Fig. 4). While helix H12 is highly conserved among most arthropods and chordates, sequences of H12 and of the loop H11-H12 are variable in Diptera and Lepidoptera. Most of the differences are conservative. The loop L5-s1, connecting helix H5 and the β-strand s1 is longer in Diptera and Lepidoptera USP-RXR, with little conservation in the additional residues (Fig. 4). Unfortunately, this region could not be modeled because its conformation is not ordered in the crystal (Billas *et al.*, 2001 ; Clayton *et al.*, 2001). Thus, further experiments are needed to decipher the putative role of this intriguing insertion.

**Divergent domains in ECR**

Despite an increase in evolutionary rates (Table 3) the ECR LBDs are rather well conserved in length and sequence (Table 2 and data not shown). This conservation enabled Wurtz *et al.* (2000) to identify the canonical 11 helices and to model 20-hydroxyecdysone binding for the Diptera *Chironomus tentans*. Thus, contrary to USP-RXR, there is no obvious divergence of the structure of ECR LBD in Diptera and Lepidoptera. This could be due to the constraint on all ECRs to presumably bind 20-hydroxyecdysone (Riddiford *et al.,* 2001).

The DBD of ECR contains six amino-acid differences specific for the Diptera-Lepidoptera group (Fig. 5A). By contrast, USP-RXR DBD sequences do not show any specific differences (Fig. 5B). Among the six differences observed for ECR, only one is not conservative and is located just upstream of the second zinc-finger. It is a hydrophobic residue in Diptera (cysteine) and Lepidoptera (isoleucine), but a polar amino-acid (glutamine) in other arthropods. Interestingly, four of these substitutions are located in or near the second zinc finger, a region known to form a dimerization interface for some nuclear receptors (Luisi *et al.*, 1991; Schwabe *et al.*, 1993).

A surprising originality of Diptera-Lepidoptera ECRs is the presence of a carboxy-terminal F domain (Fig. 6). This domain of variable length (226 in *Drosophila* and 18 in *Choristoneura*) does not show any sequence conservation between species. Other insect and arthropod ECRs possess only two to four amino-acids in carboxy-terminal of the putative helix H12 which ends the LBD. Most nuclear receptors do not contain any sizable region carboxy terminal of the LBD, including mammalian LXR and FXR, other proteins from the ECR group. Therefore, it appears that the presence of an F domain in ECR is an evolutionary acquisition of Diptera and Lepidoptera.

## DISCUSSION

This article is the first comprehensive evolutionary analysis of the ecdysone receptor, a major regulatory factor of insect development. Both partners of the ECR/USP-RXR heterodimer, which constitutes the functional ecdysone receptor, experienced a strong acceleration of evolutionary rate in Diptera and Lepidoptera. This acceleration defines a clear separation within holometabolous insects. Diptera and Lepidoptera belong to the clade Panorpida (Kristensen, 1981), with Hymenoptera as a sister group. Panorpida also includes: Trichoptera (caddisflies), the sister group of Lepidoptera, and Mecoptera (scorpionflies) and Siphonaptera (fleas) which are more closely related to Diptera (Fig. 1). The phylogenetic position of Strepsiptera is unclear (Whiting *et al.*, 1997; Rokas *et al.*, 1999). Thus the hypothesis of a unique event of acceleration in the ancestor of Diptera and Lepidoptera could be tested by isolation of ECR and USP-RXR sequences from other Panorpida. This event of acceleration could be responsible for the accelerated evolutionary rates at the base of and within these groups. We already know that USP-RXR and ECR sequences from a flea (M. Palmer, personal communication) are more similar to Diptera and Lepidoptera than to other insects (data not shown), which supports our hypothesis. Regarding evolution of metamorphosis, the divergence of the ecdysone receptor does not correlate with the different types of insect's metamorphosis. It may be necessary to isolate more full-length sequences from several species outside Panorpida to decipher a specific trend at this level.

We have identified several protein domains for which sequence divergence is specific to Diptera and Lepidoptera. All members of the nuclear hormone receptor family share the canonical LBD structure with 11 helices (H1, H3-H12) connected by loops and two short β-strands (s1 and s2). The typical activation of nuclear receptor implies the binding of the agonist ligand in the pocket. This binding triggers a repositioning of helix H12 that provides the surface for co-activator interaction and thereby allows the transactivation activity of the nuclear receptor. In the case of an antagonist, helix H12 moves precisely into the hydrophobic furrow where the co-activator interacts in the agonist conformation (Moras and Gronemeyer, 1998). In the *Drosophila* and *Heliothis* USP-RXR structures, the loop between helices H1 and H3 is located inside the hydrophobic furrow of the LBD, thereby preventing the repositioning of helix H12 and interactions with coactivators, and locking these USP-RXRs in an unusual antagonist conformation (Billas *et al.*, 2001 ; Clayton *et al.*, 2001). In the light of these results, our observation of Diptera and Lepidoptera specific sequence diversity in both the loop H1-H3 and the helix H12 suggests a form of concerted evolution between these two interacting regions of the USP-RXR LBD. This evolution may have changed the ligand-dependent transactivation activity of the protein. It may also have had an effect on the ligand binding activity, since the loop H1-H3 contains residues that interact with the phospholipid cocrystallized with *Drosophila* and *Heliothis* LBD. On the other hand, given the very strong conservation of Helix H10, it is likely that the dimerisation activity of USP-RXR LBD remained unchanged during evolution.

It is intriguing that the LBD of ECR underwent a significant increase of substitution rate in Diptera and Lepidoptera, while its structure remained apparently largely unchanged. In all insects, and presumably in all arthropods, ECR LBD binds 20-hydroxyecdysone (Riddiford *et al.,* 2001). This fundamental interaction may represent the primary selective constraint acting on this domain. However, nuclear receptor LBDs are also involved in heterodimerisation activity. This rapid evolution of ECR can be explained by adaptation to the extremely divergent USP-RXR, and eventually acquisition of new partners. It may be that the stability of the heterodimer required compensatory changes in ECR and USP-RXR, suggestive of coevolution. The differences seen in ECR DBD also suggest functional changes in dimerisation. Indeed, four of the six substitutions which are conserved among Diptera and Lepidoptera are located at positions known to be involved in protein dimerisation but not in DNA contact or nuclear localization signal (Khorasanizadeh and Rastinejad, 2001; Black *et al.*, 2001). Another difference specific to Diptera and Lepidoptera is the presence of a carboxy-terminal F domain. This difference is interesting, since it is known that when present (ERα, HNF-4) the F domain of nuclear receptors can regulate different functions of the LBD. For example, the F domain of human estrogen receptor ERα can modulate transcriptional activity and dimerisation signal, probably through interaction with the AF-2 domain (Montano *et al.*, 1995 ; Nichols *et al.*, 1998 ; Peters and Khan, 1999).

An important conclusion of this sequence analysis is that the major structural differences of USP-RXR and ECR are specific to Diptera and Lepidoptera. We hypothesize that these differences changed two functional properties of the heterodimeric ecdysone receptor during insect evolution, namely the ligand-dependent transactivation and the hetero-dimerisation activities of both USP-RXR and ECR. These hypotheses could now be tested by a comparative genetic approach using *Drosophila melanogaster* and another holometabolous insect chosen outside the Panorpida group. This should help to usefully extend our knowledge concerning the biological role of ecdysone. Indeed, our work indicates that the current model organisms used to analyze the ecdysone pathway are in fact very derived species. Therefore, extreme care must be taken when results obtained from Panorpida are generalized, notably concerning both fundamental aspects of insect development and the design of specific insecticides.

**ACKNOWLEDGMENTS**

We thank Laure Debure for samples of *Periplaneta americana*, Guillaume Balavoine for *Folsomia candida*, Frédéric Fleury for *Leptopilina heterotoma*, Elisabeth Rull for help in RT-PCR experiments, Melanie Palmer and two reviewers for their very helpful comments. CNRS, ENS, ARC Région Rhône-Alpes and MENRT funded this work.

## FIGURE LEGENDS

**Figure 1: Phylogenetic relationships between the species studied in this article.** This consensus tree is based on classical taxonomic data, as well as more specific references concerning the following groups: Diptera (Yeates and Wiegmann, 1999), Lepidoptera (Weller *et al.*, 1992; Regier *et al.*, 2001), Insects (Kristensen 1981; Whiting *et al.*, 1997) and Arthropods (Hwang *et al.*, 2001; Giribet *et al.*, 2001). Species names underlined indicate that both ECR and USP-RXR sequences are available for these species. Regarding evolution of these proteins, two artificial groups are indicated: "other insects" for all insects excluding Panorpida, and "other arthropods" for all arthropods excluding insects.

**Figure 2: Phylogenetic trees of LBD domains.** (A) ECR, (B) USP-RXR. Trees were constructed with the Neighbour-Joining method performed with the full-length LBD of ECR (17 sequences) or USP-RXR (36 sequences). Positions with a gap were excluded from the computation, resulting in 221 complete sites for ECR and 145 complete sites for USP-RXR. The RXR protein from the jellyfish *Tripedalia cystophora* was used as an outgroup to USP-RXRs, and all mammalian LXR and FXR sequences to ECRs. For legibility, outgroups are not shown. Figures at nodes are bootstrap proportions out of 1000 replicates; only values ≥ 50% are shown. The boxed bootstrap values highlight two important nodes leading to Panorpida and "other insects". Branch lengths are proportional to sequence divergence; the measure bar represents 0.1 differences per site. Diptera and Lepidoptera species are in bold.

**Figure 3: Pre-defined trees with evolutionary distances for ECR and USP-RXR**. ECR DBD (A), ECR LBD (B), USP-RXR DBD (C), USP-RXR LBD (D). Evolutionary distances between sequences were mapped on a pre-defined species consensus tree (see figure 1) using

Tree-Puzzle (Schmidt *et al.*, 2002), with the JTT substitution model (Jones *et al.*, 1992) plus rate heterogeneity between sites, estimated by a gamma law with eight categories. Branch lengths are proportional to evolutionary change; the measure bar represents 0.1 substitutions per site. Diptera and Lepidoptera species are in bold.

**Figure 4: Sequence alignment of USP-RXR LBD domains.** Sequences are aligned with human RXRα; names of Diptera species are in bold and underlined, names of Lepidoptera species are in bold. The 11 helices and the two β-strands (s1, s2) are boxed. Residues interacting with the ligand in the Ligand Binding Pocket (LBP) are indicated by asterisks (*) below the alignments. Structural data are from the following sources: Human RXRα (Bourguet *et al.*, 1995 ; Egea *et al.*, 2000), *Heliothis* USP-RXR (Billas *et al.*, 2001) and *Drosophila* USP-RXR (Clayton *et al.*, 2001). Note that helices H3 and H12 are shorter in Human RXRα, as indicated by a dashed vertical line in the amino-terminal of these helices. The gray box in the loop H1-H3 highlights a region conserved between Diptera and Lepidoptera. RT-PCR clones of five species (*Lithobius, Folsomia, Periplaneta, Alfalfa, Leptopilina*) lack some of the carboxy-terminal regions: an X indicates the end of these partial sequences. The few residues (3 to 12) following the helix H12, and therefore outside of the structurally defined LBD, are also shown on this figure.

**Figure 5: Sequences alignment of DBD domains.** (A) ECR and (B) USP-RXR. ECR sequences are aligned with *Celuca* ECR, USP-RXRs are aligned with human RXRα. Structural data are from Lee *et al.* (1993) and Khorasanizadeh and Rastinejad (2001). The two zinc-fingers are underlined on each sequence of reference and they are also indicated below the alignments; critical cysteine residues of the zinc-fingers are identified with asterisks. Names of Diptera species are in bold and underlined; names of Lepidoptera species are in bold. The gray boxes indicate ECR divergent positions between Diptera-Lepidoptera and other arthropods.

**Figure 6: Sequence alignment of ECR F domains.** Sequences are aligned with *Drosophila* ECR. Names of Diptera species are in bold and underlined; names of Lepidoptera species are in bold. Putative helix H12 (Wurtz *et al.*, 2000) is boxed. Amino-acids following the helix H12 are numbered above *Drosophila* sequence. The total number of residues in carboxy-terminal domains of ECR proteins is indicated at the end of each sequence.

**Table1: Accession number and phylogenetic origin of proteins used in this study.**

[a] Partial sequences. [b] This paper.

| Group | Species | USP-RXR | ECR (and related) |
|---|---|---|---|
| Diptera | *Drosophila melanogaster* | P20153 | P34021 |
| | *Ceratitis capitata* | | CAA11907 |
| | *Lucilia cuprina* | | O18531 |
| | *Calliphora vicina* | | AAG46050 |
| | *Sarcophaga crassipalpis* | AAF44674 [a] | AAF44673 [a] |
| | *Aedes aegypti* | AAG24886 | P49880 |
| | *Aedes albopictus* | AAF19033 | AAF19032 |
| | *Chironomus tentans* | AAC03056 | P49882 |
| Lepidoptera | *Bombyx mori* | S44490 | P49881 |
| | *Manduca sexta* | P54779 | P49883 |
| | *Heliothis virescens* | 14278415 [a] | O18473 |
| | *Junonia coenia* | | CAB63485 [a] |
| | *Bicyclus anynana* | | CAB63236 [a] |
| | *Choristoneura fumiferana* | AAC31795 | AAC61596 |
| Hymenoptera | *Apis mellifera* | AAF73057 | |
| | *Leptopilina heterotoma* | AY157931 [ab] | AY157932 [ab] |
| Coleoptera | *Tenebrio molitor* | CAB75361 | CAA72296 |
| | *Alfalfa weezi* | AY157933 [ab] | |
| Orthoptera | *Locusta migratoria* | AAF00981 | AAD19828 |
| Dictyoptera | *Periplaneta americana* | AY157928 [ab] | |
| Collembola | *Folsomia candida* | AY157930 [ab] | |
| Crustacea | *Celuca pugilator* | AAC32789 | AAC33432 |
| Chelicerata | *Amblyomma americanum* | RXR1: AAC15588 | AAB94566 |
| | | RXR2: AAC15589 | |
| Myriapoda | *Lithobius forficatus* | AY157929 [ab] | |
| Vertebrata | *Homo sapiens* | RXRα: CAA36982 | LXRα: Q13133 |
| | | RXRß: AAA60293 | LXRß: P55055 |
| | | RXRγ: AAA80681 | FXR: AAB08107 |
| | *Rattus norvegicus* | RXRα: AAA42093 | LXRα: Q62685 |
| | | | LXRß: Q62755 |
| | | | FXR: A56918 |
| | *Rattus rattus* | RXRß: AAA42025 | |
| | *Mus musculus* | RXRα: AAA40080 | LXRα: Q9Z0Y9 |
| | | RXRß: CAA46963 | LXRß: Q60644 |
| | | RXRγ: CAA46964 | FXR: NP_033134 |
| | *Gallus gallus* | RXRγ: CAA41743 | |
| | *Xenopus laevis* | RXRα: P51128 | |
| | | RXRß: S73269 | |
| | | RXRγ: P51129 | |
| | *Danio rerio* | RXRα: AAC59719 | |
| | | RXRß1: AAC59722 | |
| | | RXRß2: AAC59721 | |
| | | RXRγ: AAC59720 | |
| Urochordata | *Polyandrocarpa misakiensis* | BAA82618 [a] | |
| Cnidaria | *Tripedalia cystophora* | AAC80008 | |

**Table 2. Average identity percentages of pairwise comparisons for DBD and LBD domains of USP-RXR and ECR.** #: no Chordate ECR sequences to compare.

| Groups of species | | DBD | | LBD | |
|---|---|---|---|---|---|
| | | USP-RXR | ECR | USP-RXR | ECR |
| Diptera-Lepidoptera > | Other Insects | 94.6 ± 1,5 | 88.75 ± 1,1 | 49.1 ± 3,1 | 64.4 ± 3,1 |
| | Other Arthropods | 93.6 ± 2 | 88.8 ± 1,8 | 43.65 ± 2,1 | 58.2 ± 3,2 |
| | Chordates | 82.7 ± 1,8 | # | 46.7 ± 2 | # |
| Other Insects > | Other Arthropods | 93.3 ± 2,1 | 96.9 ± 0,7 | 68 ± 4,6 | 67.7 ± 2,7 |
| | Chordates | 84.4 ± 1,7 | # | 69.8 ± 3,3 | # |
| Other Arthropods > | Chordates | 82.4 ± 1,5 | # | 70.1 ± 1,9 | # |

**Table 3. Comparison of evolutionary rates for USP-RXR and ECR LBDs between three groups of arthropods.** Values idicated are: substitution rate difference ± standard deviation. The probability associated to the test is indicated as follows: Not Significant *(NS)* > 5%; * ≤ 5%; ** ≤ 1%; *** ≤ 0.5%.

| **Groups of species** | | **USP-RXR LBD** | **ECR LBD** |
|---|---|---|---|
| Diptera-Lepidoptera > | Other Insects | 0.307 ± 0.068 *** | 0.122 ± 0.054 * |
| Diptera-Lepidoptera > | Other Arthropods | 0.365 ± 0.082 *** | 0.129 ± 0.057 * |
| Other Insects > | Other Arthropods | 0.0574 ± 0.041 *NS* | 0.0064 ± 0.049 *NS* |

**Figure 1**

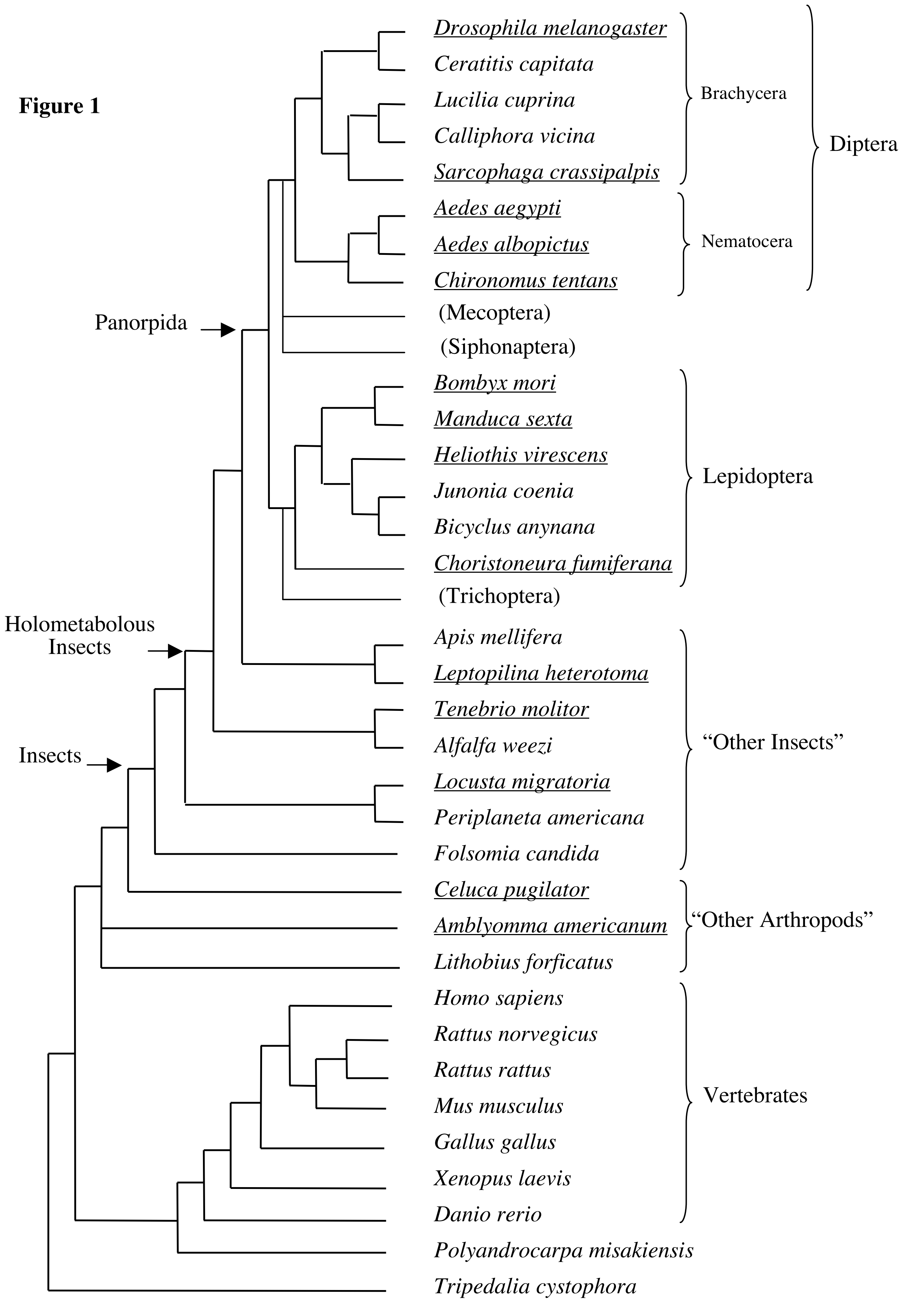

FIG 2A
ECR LBD

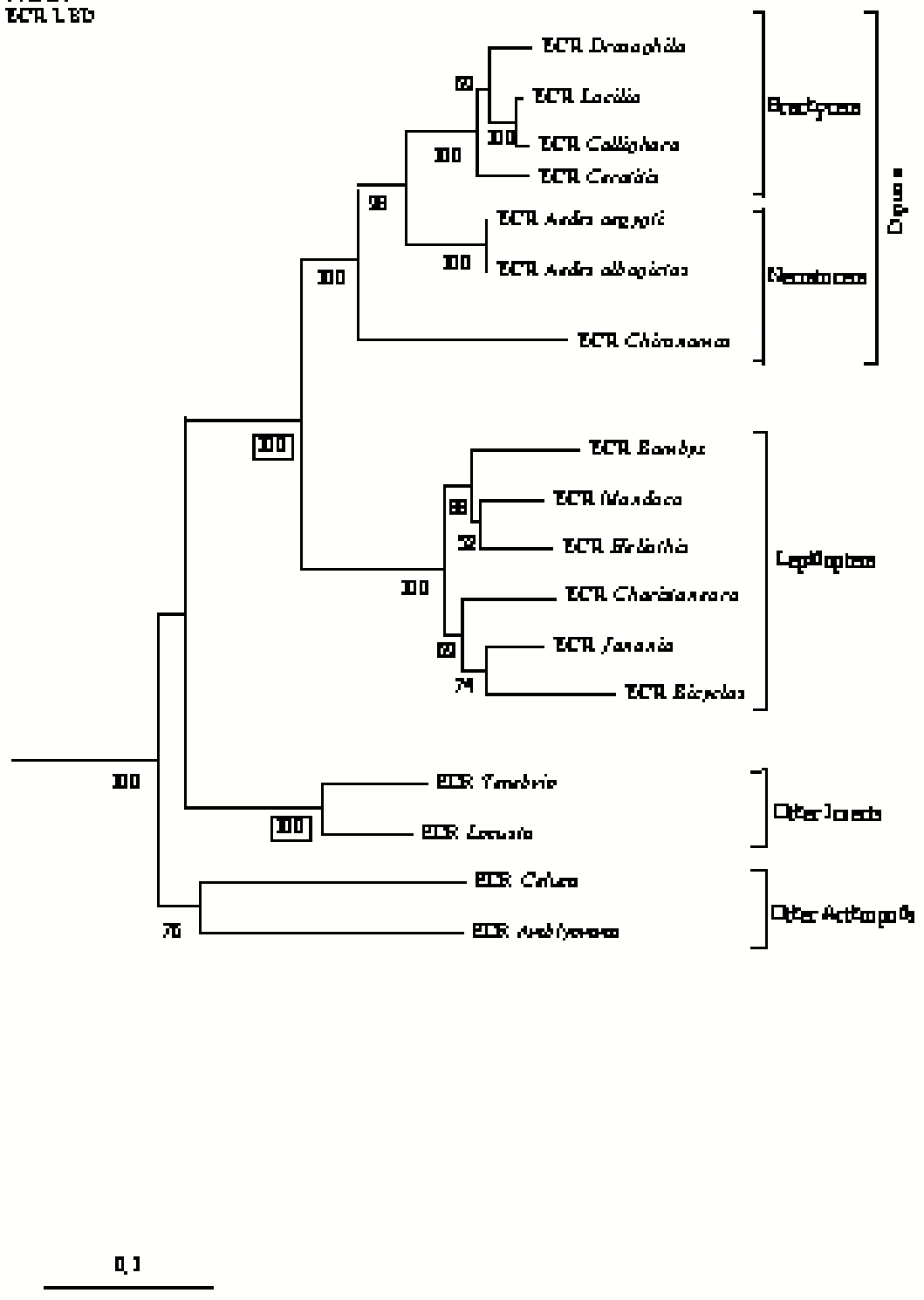

FIG 2B
USP-RXR LBD

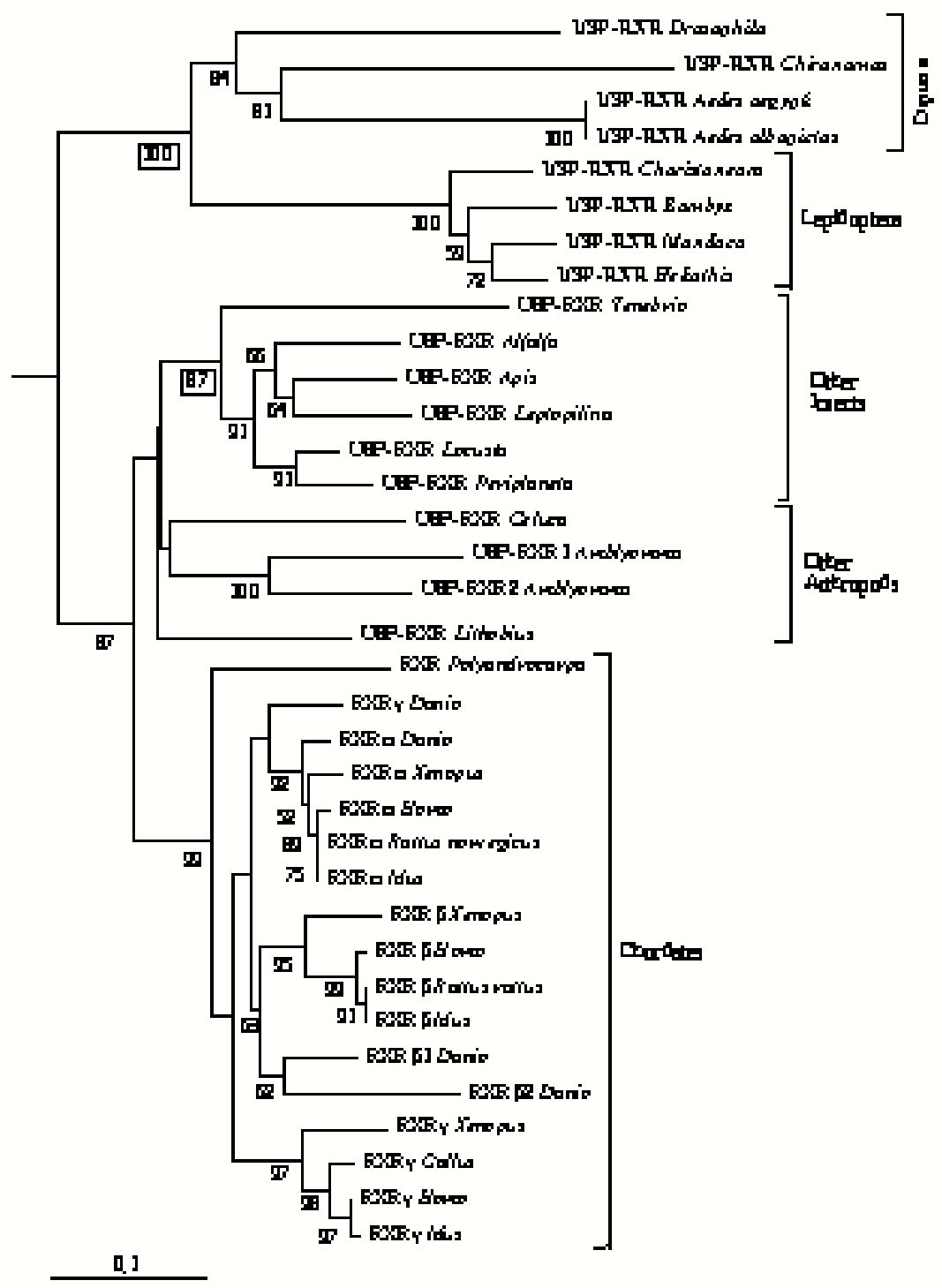

**FIG 3A**
**ECR DBD**

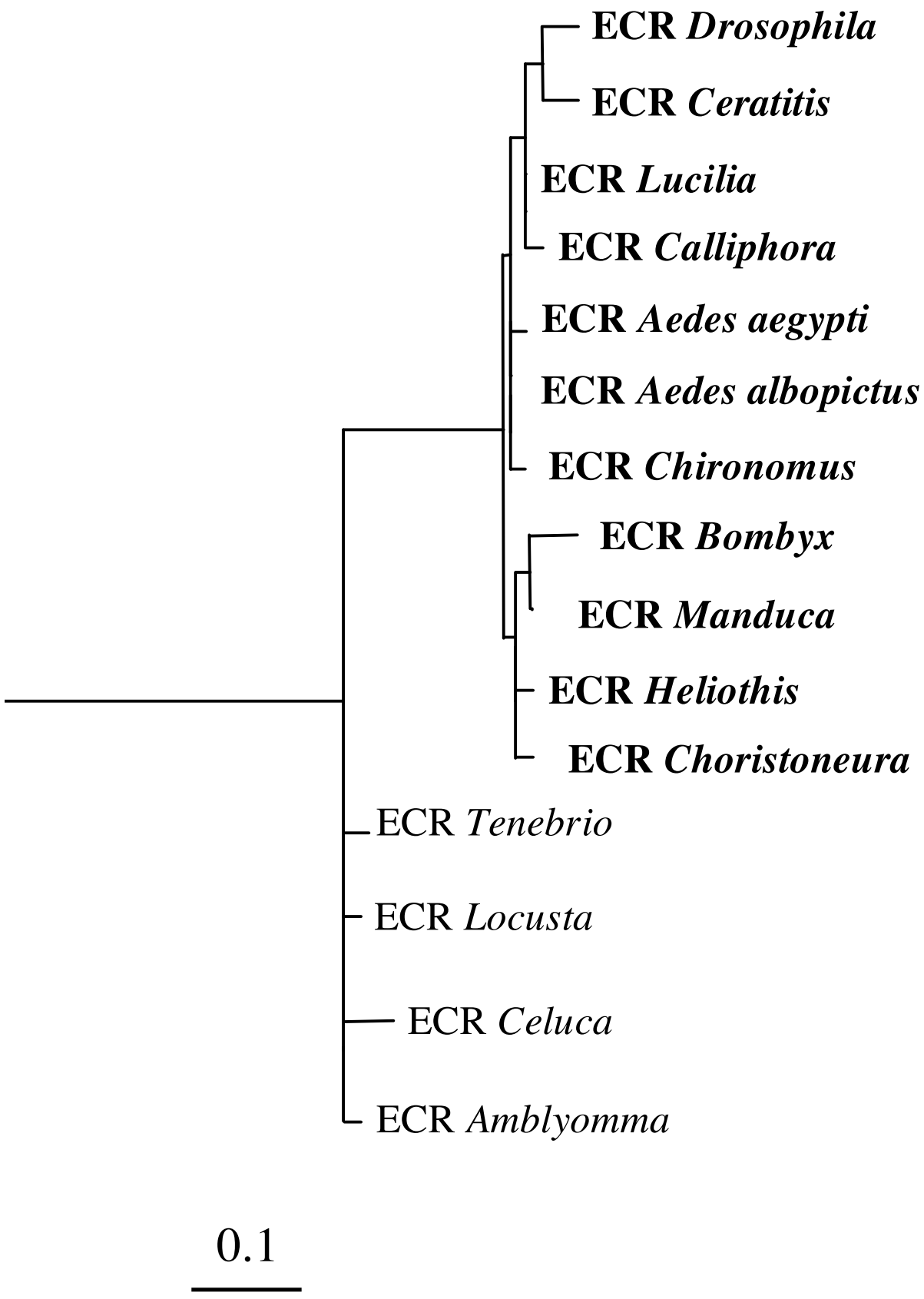

**FIG 3B**
**ECR LBD**

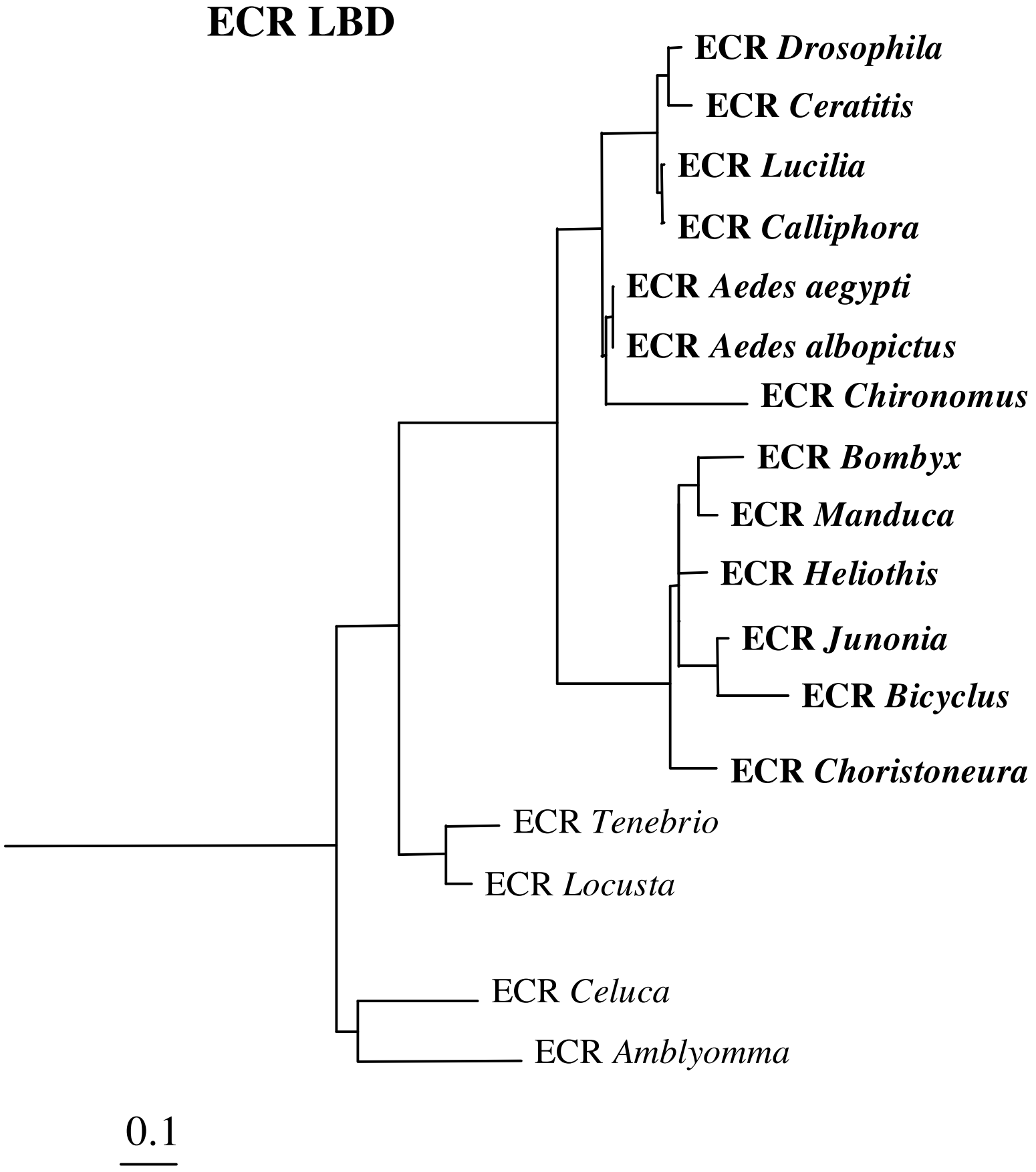

**FIG 3C**
**USP-RXR DBD**

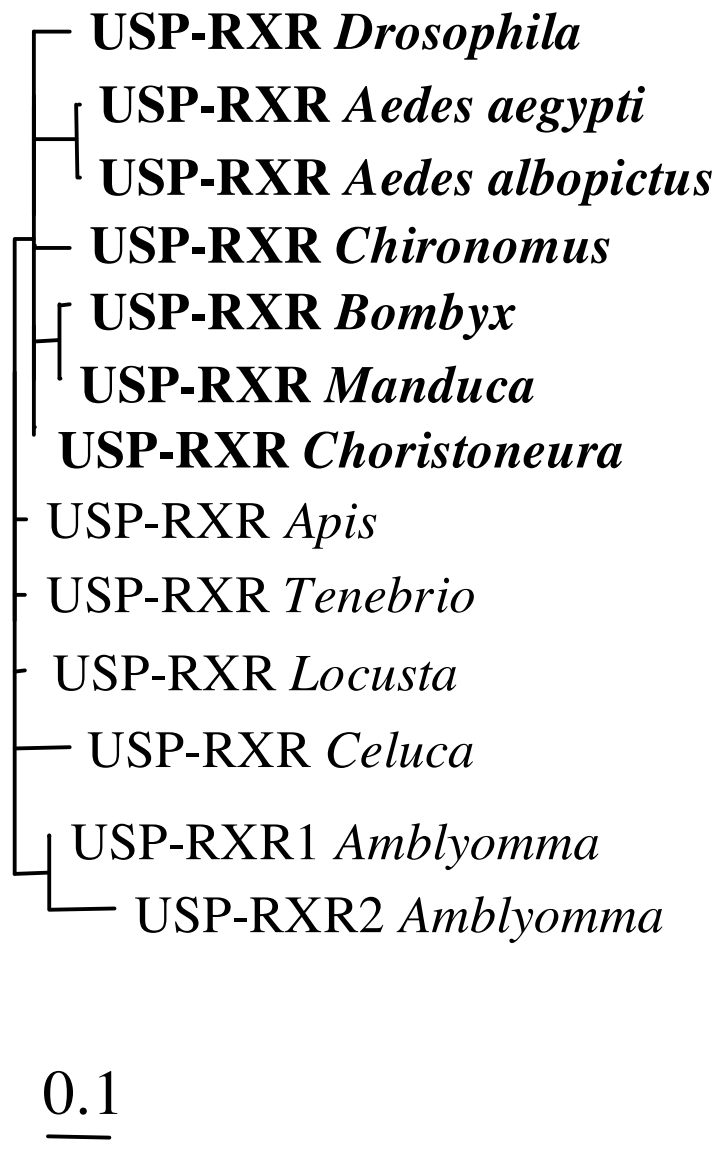

**FIG 3D**
**USP-RXR LBD**

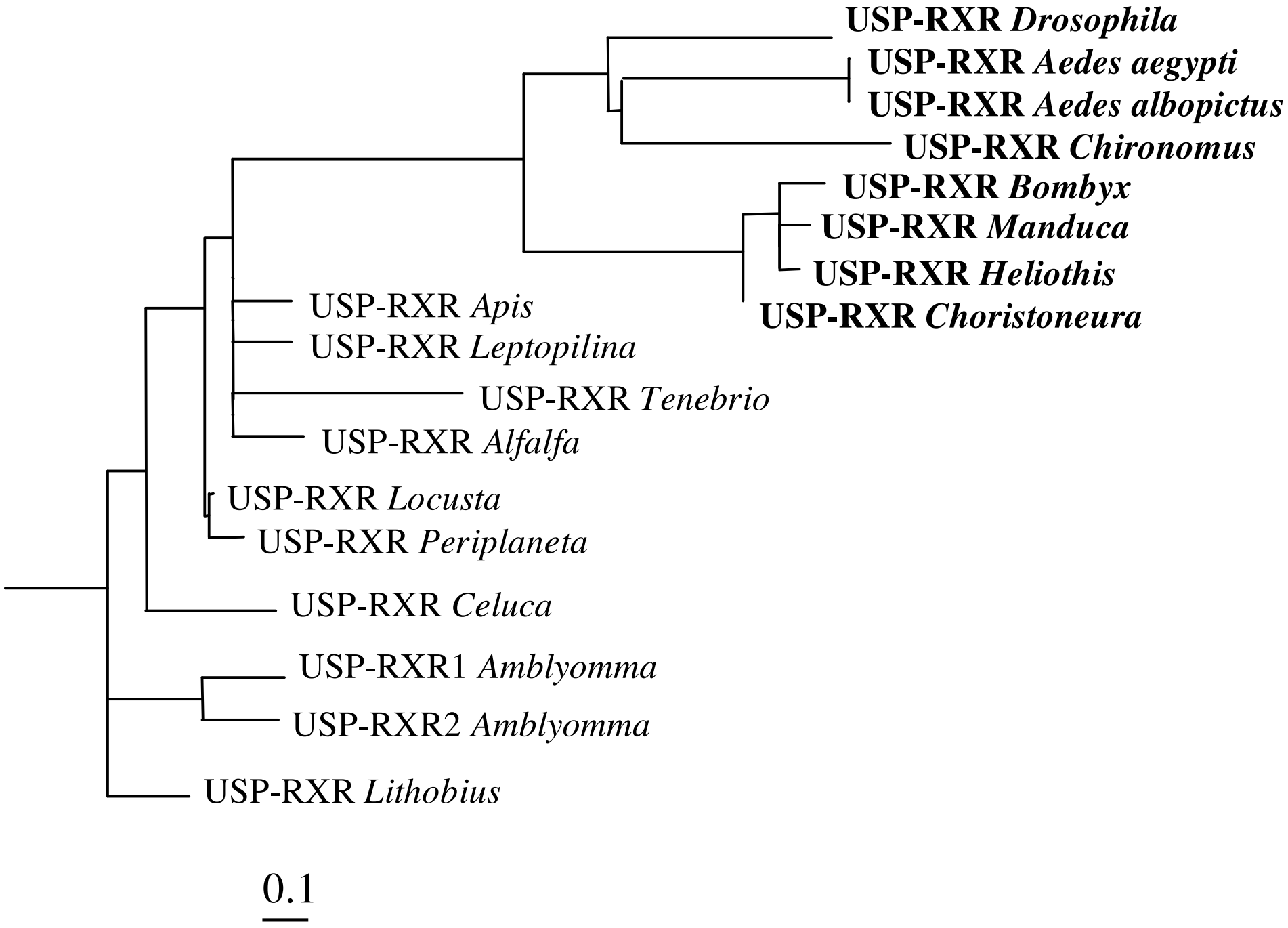

FIG 4

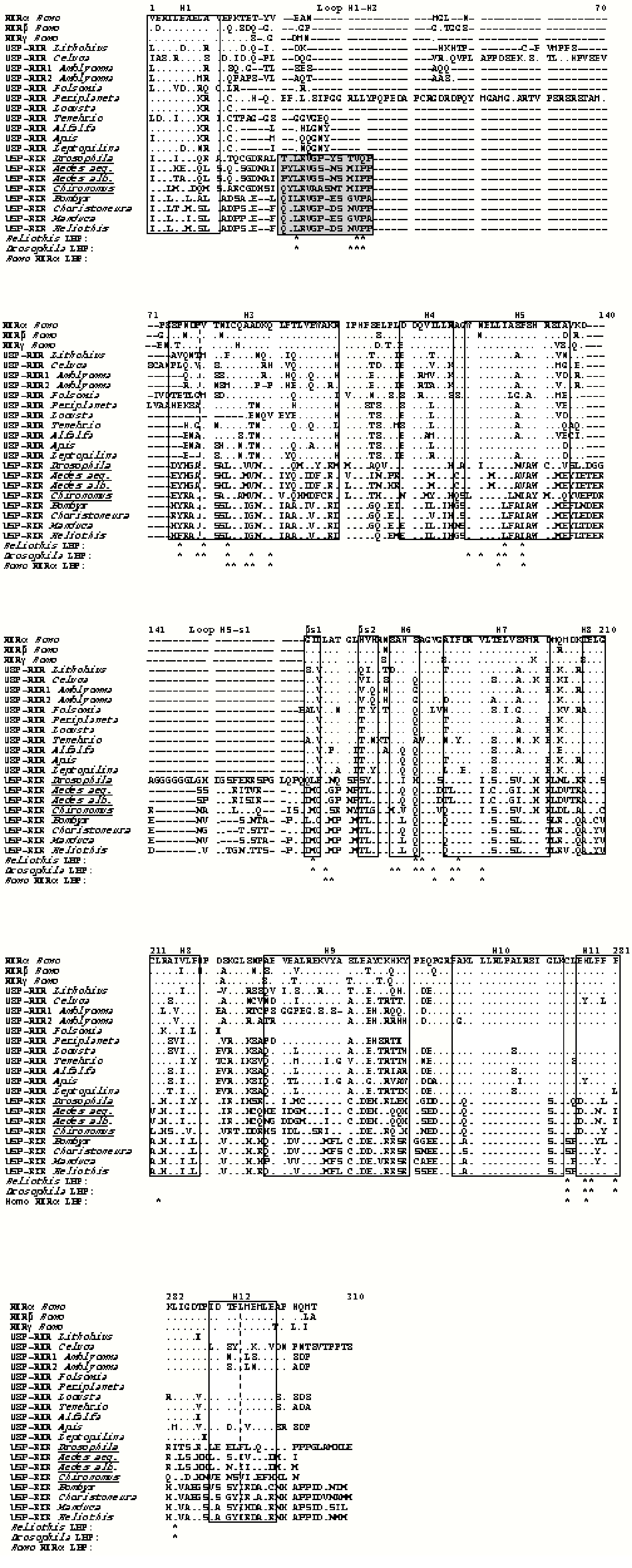

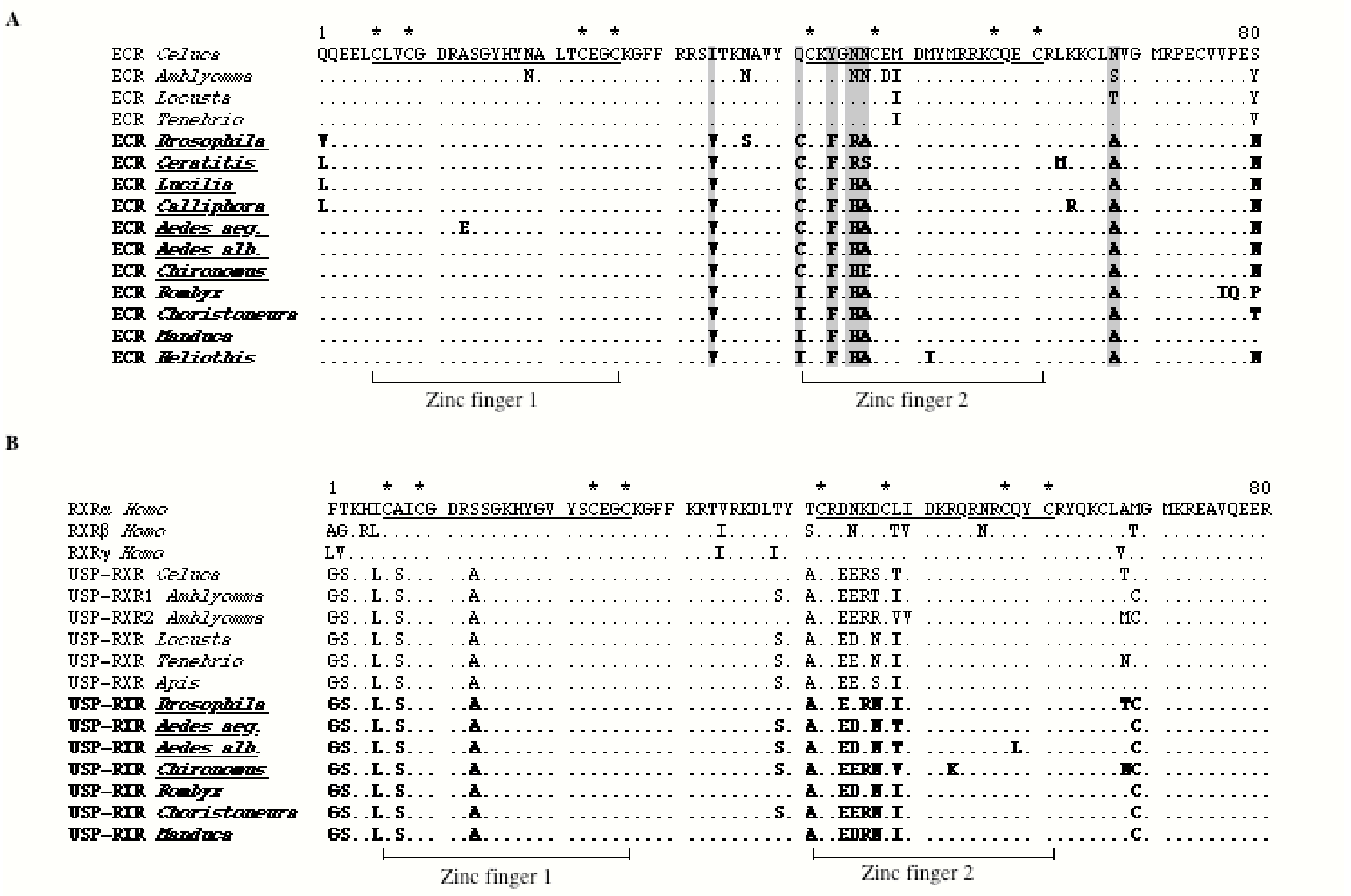
A
ECR Celuca QQEELCLVCG DRASGYHYNA LTCEGCKGFF RRSITKNAVY QCKYGNNCEM DMYMRRKCQE CRLKKCLNVG MRPECVVPES
ECR Amblyomma
ECR Locusta
ECR Tenebrio
ECR Drosophila
ECR Ceratitis
ECR Lucilia
ECR Calliphora
ECR Aedes aeg.
ECR Aedes alb.
ECR Chironomus
ECR Bombyx
ECR Choristoneura
ECR Manduca
ECR Heliothis
Zinc finger 1
Zinc finger 2
B
RXRα Homo FTKHICAICG DRSSGKHYGV YSCEGCKGFF KRTVRKDLTY TCRDNKDCLI DKRQRNRCQY CRYQKCLAMG MKREAVQEER
RXRβ Homo
RXRγ Homo
USP-RXR Celuca
USP-RXR1 Amblyomma
USP-RXR2 Amblyomma
USP-RXR Locusta
USP-RXR Tenebrio
USP-RXR Apis
USP-RXR Drosophila
USP-RXR Aedes aeg.
USP-RXR Aedes alb.
USP-RXR Chironomus
USP-RXR Bombyx
USP-RXR Choristoneura
USP-RXR Manduca
Zinc finger 1
Zinc finger 2

FIG 5

```
                      H12      1                                                                      61
ECR Drosophila     KFLEEIWDV H AIPPSVQSHL QITQEEHERL ERAERMRASV GGAITAGIDC DS----------A STSAAAAAAQ   ->226
ECR Ceratitis      ......... . .V.......I .A..A.R.G. .PNTAVATTS TS.AASSSPV RP----------L TW             46
ECR Lucilia        ......... . .........I .A..A.KAGP GSSGNNIG-- ---HFSSRHL IF----------L HKYLDGNIIL   -> 82
ECR Calliphora     ......... . .........I .A..A.KAGP GSPGYHFG.. AHGHFS.RHL IV----------L HKYLDGDIIL   -> 94
ECR Aedes aeg.     R........ Q D....M.AQM HSHGTQ--SS SSSSSSSS.S SNGSSN.NSS SNSNSSQHGP HPHPHGQQLT   ->105
ECR Aedes alb.     R........ Q D....M.AQM HSHGTPQSSS SSSSSSSS.S SNGSSN.NSN SN--------.P HPHPHGQQLT   ->110
ECR Chironomus     R....V... G DVNNQTTATT NTENIV...I N.N                                              24
ECR Bombyx         P........ A EVATTHPT-- VLPPTNPVV.                                                  19
ECR Choristoneura  P........ A DMSHTQPPP- ILESPTNL                                                    27
ECR Manduca        P........ A EVSTTQPTPG VAA.VTPIVV DNPAAL                                           18
ECR Heliothis      P........ A DVATTATPV- AAEAPAPLAP APPA.PP.T.                                       30
ECR Tenebrio       P..D..... D LKA                                                                     4
ECR Locusta        P..A..... I P                                                                       2
ECR Amblyomma      P..A....I Q E                                                                       2
ECR Celuca         P..A..... S GY                                                                      3
```

FIG 6